\documentclass[conference,letterpaper]{IEEEtran}

\usepackage[
  letterpaper,
  left=0.64in,
  right=0.64in,
  top=0.76in,
  bottom=1.05in
]{geometry}

\usepackage[utf8]{inputenc} 
\usepackage[T1]{fontenc}
\usepackage{url}
\usepackage{ifthen}
\usepackage{cite}
\usepackage[cmex10]{amsmath} % Use the [cmex10] option to ensure complicance
\usepackage{amsfonts}
\usepackage{amssymb}
\usepackage{amsthm}
\usepackage{enumitem}
\usepackage{xcolor}

\newtheorem{theorem}{Theorem}

\newtheorem{definition}{Definition}

\newtheorem{proposition}{Proposition}

\theoremstyle{definition}
\newtheorem{remark}{Remark}

\theoremstyle{definition} \newtheorem{example}{Example}

\def \extended {1}

\begin{document}
%\title{From Submodularity to Matrix Determinants: Strengthening Han's, Sz\'asz's, and Fischer's Inequalities}
\title{Conditioning Strengthens Classical Entropy and Matrix Inequalities: A Submodular Approach}
\author{%
  \IEEEauthorblockN{Gunank Jakhar, Gowtham R. Kurri, Suryajith Chillara}
  \IEEEauthorblockA{International Institute of Information Technology, Hyderabad \\
                    Hyderabad, India.\\ 
                    Email: \{gunank.jakhar@research., gowtham.kurri@, suryajith.chillara@\}iiit.ac.in}
  \and
  \IEEEauthorblockN{Vinod M. Prabhakaran}
  \IEEEauthorblockA{Tata Institute of Fundamental Research\\
                    Mumbai, India.\\
                    Email: vinodmp@tifr.res.in}
}

\maketitle

\begin{abstract}
\if \extended 0%
THIS PAPER IS ELIGIBLE FOR THE STUDENT PAPER AWARD.
\fi
Dembo, Cover, and Thomas (1991) developed an elegant information-theoretic framework for proving determinantal inequalities for positive definite matrices, which relies on the structural inequalities of differential entropy. Submodular functions, which subsume entropy, inherently satisfy these structural inequalities because they obey generalized forms of the fundamental properties of entropy---a chain rule and the property that conditioning reduces the function's value (under an appropriate definition of conditioning). Within this framework, applying subadditivity, Han's inequality (1978), and partition subadditivity (i.e., subadditivity over a partition) yields Hadamard's, Sz\'asz's, and Fischer's inequalities, respectively. Furthermore, using conditional subadditivity, this framework recovers Ky Fan's inequality (1955), a strengthening of Hadamard's inequality. We highlight a previously unnoticed information-theoretic insight behind this improvement: conditioning yields a tighter upper bound on the joint entropy than the unconditional one.

% This improvement fundamentally arises because conditional subadditivity yields a tighter upper bound on the joint entropy than the one obtained via unconditional subadditivity. 

In this paper, we establish conditional strengthenings of Han's inequality and partition subadditivity in the general setting of submodular functions. We derive equality conditions for these strengthened bounds and characterize when they strictly improve their unconditional counterparts. We specialize these results to differential entropy and apply them to establish strengthened versions of Sz\'asz's and Fischer's inequalities. The strengthening of Sz\'asz's inequality recovers Ky Fan's inequality as a special case, and is strictly stronger than the classical Sz\'asz's inequality for any non-diagonal positive definite matrix. We also derive an inequality concerning eigenvalues, which generalizes and strictly strengthens a corresponding eigenvalue inequality of Ky Fan. We provide numerical examples to explicitly illustrate the strengthening achieved by our proposed matrix determinantal bounds. 
\end{abstract}

\section{Introduction}
A set function $f: 2^{[1:n]} \rightarrow \mathbb{R}$ is said to be submodular if it satisfies the diminishing returns property; that is, the marginal value of adding an element to a set decreases as the set grows larger~\cite{Fujishige05}. Submodular functions have applications across various fields, including machine learning~\cite{krause2007near}, combinatorial optimization~\cite{ConfortiC84}, algorithmic game theory~\cite{lehmann2001combinatorial}, social networks~\cite{kempe2003maximizing}, and statistical physics~\cite{VanEnterFS93}. An example of a submodular function is entropy $H(\cdot)$ (as well as differential entropy $h(\cdot)$)---for jointly distributed random variables $X_1,\dots,X_n$, the function $f(S) = h(X_S)$ for $S\subseteq [1:n]$ is submodular~\cite{Fujishige05}. Other examples include graph-cut functions~\cite{Bach13}, set cover~\cite{Bach13}, matroid rank functions~\cite{Fujishige05}, and facility location~\cite{IyerKBA22}.

Entropy satisfies many structural inequalities concerning subsets of random variables. Notably, Han's inequality~\cite{Han78} shows that the joint entropy is upper bounded by the weighted average of the entropies of all $k$-sized subsets of the random variables, for any $k \in [1:n]$. Further, partition subadditivity states that the joint entropy of a collection of random variables is upper bounded by the sum of the entropies of subsets forming a partition of that collection. These inequalities have found applications across diverse fields, including extremal graph theory~\cite{BoucheronLM13,Sason22}, communication complexity~\cite{BarYossefJKS02}, private information retrieval~\cite{SunJ18}, geometric inequalities~\cite{CarlenC09}, and coin-weighing problems~\cite{Pippenger99}. General submodular functions inherently satisfy these two inequalities because they obey the fundamental properties used to prove those inequalities: a chain rule and the property that conditioning reduces the function's value (under an appropriate definition of conditioning)~\cite{Fujishige05,MadimanT10}.

An elegant information-theoretic framework for proving determinantal inequalities for positive definite matrices was developed by Cover and El Gamal~\cite{cover1983information} and Dembo~\emph{et al.}\cite{DemboCT91}. For instance, the standard subadditivity of differential entropy over a partition of singletons yields Hadamard's inequality~\cite{HornJ12}. Furthermore, applying Han's inequality for $k$-sized subset families and partition subadditivity yields Sz\'asz's and Fischer's inequalities~\cite{HornJ12}, respectively. Within this same framework, Dembo~\emph{et al.}~\cite{DemboCT91} showed that applying conditional subadditivity of differential entropy recovers Ky Fan's inequality~\cite{Fan55}, originally introduced as a strengthening of Hadamard's inequality. 
% This strengthening implicitly arises from the simple fact that the upper bound on $h(X_{[1:n]})$ obtained from conditional subadditivity applied to $h(X_{[1:p]}|X_{[p+1:n]})$ is tighter than the one obtained from standard subadditivity; that is, 
% \begin{align}\sum_{i=1}^p h(X_i|X_{[p+1:n]}) + h(X_{[p+1:n]}) \leq \sum_{i=1}^n h(X_i).
% \end{align}
% However, to the best of our knowledge, analogous strengthenings of Han's inequality and partition subadditivity through conditioning have not been addressed. 
However, the information-theoretic insight behind this improvement has, until now, gone unnoticed in the literature. Specifically, we highlight a simple yet fundamental observation: conditioning yields a tighter upper bound on joint entropy than the standard unconditional bound:
\begin{align}
    \sum_{i=1}^p h(X_i|X_{[p+1:n]}) + h(X_{[p+1:n]}) \leq \sum_{i=1}^n h(X_i).
\end{align}
Abstracting its underlying mechanism allows us to extend this conditioning principle beyond basic subadditivity, establishing conditional strengthenings of classical inequalities. In particular, our main contributions are as follows:

%Our main contributions are as follows.
\begin{itemize}[leftmargin=*]
    \item We establish conditional strengthenings of Han's inequality and partition subadditivity in the broad framework of general submodular functions. We also derive equality conditions for these strengthened bounds and characterize precisely when these inequalities are strictly stronger than their unconditional counterparts (Theorems~\ref{StrongHanInequalitySubmodular} and \ref{ConditionalSubadditivitySubmodular}). 
    \item We specialize Theorems~\ref{StrongHanInequalitySubmodular} and \ref{ConditionalSubadditivitySubmodular} to differential entropy (Propositions~\ref{StrongHanInequality} and \ref{ConditionalSubadditivity}) and apply them to Gaussian random variables to obtain strengthenings of Sz\'asz's and Fischer's inequalities, respectively (Theorems~\ref{StrongSzasz} and \ref{StrongFischer}). Notably, the bound established in Theorem~\ref{StrongSzasz} is typically strictly tighter than the one in Sz\'asz's inequality for non-diagonal positive definite matrices (Remark~\ref{remark:szaszsrtong}). Theorem~\ref{StrongSzasz} also recovers Ky Fan's inequality~\cite[Theorem~1]{Fan55} as a special case, thereby yielding a previously unstated characterization of precisely when it strictly improves upon Hadamard's inequality. 
    \item Using Theorem~\ref{StrongSzasz}, we prove an inequality concerning the eigenvalues of positive definite matrices (Theorem~\ref{EigenvaluesStrongKyFan}). This generalizes and strictly strengthens the corresponding eigenvalue inequality established by Ky Fan~\cite[Theorem~4]{Fan55}. We also provide numerical examples to explicitly illustrate the strengthening for our matrix determinantal inequalities.
\end{itemize}

\section{Preliminaries}\label{Prelim}
\textit{Notation}: We use $[i : i+k]$ to represent the set $\{i, i+1, \ldots, i+k\}$, where $i,k\in\mathbb{N}$. The power set of a set $A$ is denoted by $2^A$. For $A\subseteq[1:n]$, $A^{\mathrm c}$ denotes its complement with respect to $[1:n]$. $H(X)$ and $h(X)$ denote the entropy of a discrete random variable $X$ and the differential entropy of a continuous random variable $X$, respectively. Throughout this paper, all differential entropies appearing in the statements and proofs are assumed to be finite. For jointly distributed random variables $X_1,\dots,X_n$, let $X_{S}=(X_i:i\in S)$, for $S\subseteq [1:n]$. $|K|$ is used to denote the determinant of a matrix $K$. For an $n \times n$ matrix $K$, and $S,T \subseteq [1:n]$, $K(S,T)$ denote the submatrix of $K$ corresponding to the rows and columns indexed by the elements of $S$ and $T$, respectively. We use $K(S)$ to denote the principal submatrix $K(S,S)$, for $S\subseteq [1:n]$. We adopt the convention that $|K(\emptyset)| = 1$.

\begin{definition}[Submodular Functions~\cite{Fujishige05}]
A set function $f: 2^{[1:n]} \rightarrow \mathbb{R}$ is called submodular if
\begin{align}
    f(S) + f(T) \geq f(S \cup T) + f(S \cap T),\ \forall S,T \subseteq [1:n].
\end{align}
\end{definition}

It is known that submodular functions satisfy a conditioning property and a chain rule under an appropriately defined notion of conditioning, e.g., see \cite{MadimanT10}. Specifically, for $S,T\subseteq [1:n]$, the conditional version of submodular function is defined as $f(S \mid T) = f(S\cup T) -f(T)$. Let $S,T,U$ be disjoint subsets of $[1:n]$, then \cite[Lemma~IV]{MadimanT10} states that
\begin{align}
      f(S \mid T, U) &\leq f(S \mid T), \label{eqref: conditioning}\\
       f([1:n]) &= \sum_{i=1}^n f(\{i\} \mid [1:i-1]), \label{eqref: chain_rule}
\end{align}
where $f(S \mid T,U):=f(S \mid T\cup U)$.

These two properties lead to an inequality for submodular functions in terms of families of subsets. The following theorem was originally formulated by Han \cite{Han78} specifically for entropy, which is a submodular function~\cite{Fujishige05}. The underlying subset-averaging principle, however, is fundamentally a structural property of all submodular functions, as implicitly captured by Fujishige \cite{Fujishige78}. We refer to this generalized result as Han's inequality for submodular functions.

\begin{theorem}[\!\!{\cite{Han78},\cite[Section~IV]{Fujishige78}}]\label{HanInequality}
    Let $f$ be any submodular function with $f(\emptyset) = 0$. Then, for any $1\leq k\leq n$,
    \begin{align}\label{eq: HanIneq}
        f([1:n]) \leq \frac{1}{\binom{n-1}{k-1} }  \sum\limits_{\substack{S \subseteq [1:n]: \\ |S| = k}} f(S).
    \end{align}
\end{theorem}
The special case of $k=1$ in \eqref{eq: HanIneq} can be seen as the subadditivity property of submodular functions. In particular, by taking $f(S) = h(X_S)$ for $S \subseteq [1:n]$, where $X_1, \dots, X_n$ are jointly distributed random variables, we recover the well-known subadditivity property of (differential) entropy: $h(X_{[1:n]}) \leq \sum_{i=1}^n h(X_i)$.

Furthermore, it is immediate that basic subadditivity property of submodular functions naturally extends to any partition of $[1:n]$, a property we will refer to as \textit{partition subadditivity}. That is, for any partition $\mathcal{P}$ of $[1:n]$, we have
\begin{align}\label{eq: PartitionSubadditivity}
f([1:n]) \leq \sum_{S \in \mathcal{P}} f(S).
\end{align}

Several classical inequalities for determinants of positive-definite matrices can be obtained as consequences of entropy inequalities. This information-theoretic viewpoint was systematically developed by Dembo, Cover, and Thomas~\cite{DemboCT91}. We recall some inequalities below, along with their information-theoretic proof ideas.
\begin{theorem}\label{thm:matrixineq}
    Let $K$ be a positive definite $n \times n$ matrix. Then
    \begin{itemize}[leftmargin=*]
        \item (Hadamard's Inequality~\cite{HornJ12}) $|K| \leq \prod_{i = 1}^n K_{ii}$ with equality if and only if $K$ is diagonal, where $K_{ii}$ denotes the $i$-th diagonal entry in $K$.
        \item (Sz\'asz's Inequality~\cite{HornJ12}) For $1\leq k \leq n$, 
        \begin{align}\label{eq: Szasz}
        |K| \leq \bigg(\prod\limits_{S \subseteq [1:n]:|S| = k} |K(S)|\bigg)^{\frac{1}{\binom {n-1}{k-1}}}.
        \end{align}
        For $1\leq k<n$, equality holds if and only if $K$ is diagonal.
        \item (Fischer's Inequality~\cite{HornJ12}) For any partition $\mathcal{P}$ of $[1:n]$, $|K|\leq \prod_{S \in \mathcal{P}}|K(S)|$ with equality if and only if $K(S,S') = 0$ for all distinct $S, S' \in \mathcal{P}$.
    \end{itemize}
\end{theorem}
Hadamard's inequality can be obtained by applying subadditivity of differential entropy,
\begin{align}\label{eqn:subadditivity}
    h(X_{[1:n]}) \leq \sum_{i=1}^n h(X_i),
\end{align}
to a zero-mean Gaussian random vector $(X_1,\dots,X_n)$ with covariance matrix $K$, together with the identity \(h(X_S)=\frac{|S|}{2}\ln(2\pi e)+\frac12\ln |K(S)|\). Similarly, Sz\'asz's and Fischer's inequalities follow by applying Han's inequality for differential entropy,
\begin{align}\label{eqn:Han-entropy}
      h(X_{[1:n]}) \leq \frac{1}{\binom{n-1}{k-1} }  \sum\limits_{S \subseteq [1:n]: |S| = k} h(X_S),
\end{align}
and partition subadditivity for differential entropy 
\begin{align}
    h(X_{[1:n]})\leq \sum_{S\in\mathcal{P}}h(X_S),
\end{align}
for a partition $\mathcal{P}$ of $[1:n]$, respectively, to the same Gaussian random vector.

Ky Fan~\cite{Fan55} proved the following strengthening of Hadamard's inequality. 
\begin{theorem}[\!\!{\cite[Theorem~1]{Fan55}}]\label{KyFan}
    For a positive definite $n \times n$ matrix $K$, for any positive integer $p \leq n$, and $Q = [p+1:n]$
    \begin{align}\label{eq: KyFan}
        |K| \leq |K(Q)|\prod\limits_{i = 1}^p \frac{|K(\{i\} \cup Q)|}{|K(Q)|}.
    \end{align}
\end{theorem}
The proof of \eqref{eq: KyFan}, as given in \cite{DemboCT91}, uses conditional subadditivity of entropy,
\begin{align}\label{eqn:condsubadd}
    h(X_{[1:p]}|X_{[p+1:n]})\leq \sum_{i=1}^ph(X_i|X_{[p+1:n]}),
\end{align}
applied to the same Gaussian random vector as above. The strength of \eqref{eq: KyFan} compared to Hadamard's inequality can be
seen from the simple but useful observation that the upper bound on $h(X_{[1:n]})$ obtained from conditional subadditivity in \eqref{eqn:condsubadd} is no larger than the one obtained from subadditivity in \eqref{eqn:subadditivity}. Indeed, comparing the two respective bounds directly yields
\begin{align}
     \sum_{i=1}^ph(X_i|X_{[p+1:n]})+h(X_{[p+1:n]})&\leq \!\sum_{i=1}^ph(X_i)+\!\!\sum_{i=p+1}^nh(X_i)\\
    &=\sum_{i=1}^nh(X_i)\label{subaddstrong}.
\end{align}

To the best of our knowledge, analogous strengthenings for Sz\'asz's and Fischer's inequalities through conditioning were not known prior to this work.

\section{Strengthening Han's Inequality and Partition Subadditivity for Submodular Functions}
While conditional subadditivity of entropy directly shows that conditioning yields a tighter bound on $h(X_{[1:n]})$ than standard subadditivity as seen in \eqref{subaddstrong}, extending this comparison to incorporate arbitrary families of subsets of $[1:n]$ is less immediate. Indeed, the conditioning set may intersect several members of the family, and the resulting bounds may involve different subset families and weights, so a direct term-by-term application of the conditioning property does not suffice in general. We establish this comparison in the broader combinatorial framework of submodular functions, focusing on Han’s inequality in \eqref{eq: HanIneq} for families consisting of all $k$-element subsets, as well as on partition subadditivity in \eqref{eq: PartitionSubadditivity}. 

The first inequality in the following theorem is a conditional form of Han's inequality; specifically, it is Han's inequality in \eqref{eq: HanIneq} applied to the contracted set function $g: S\subseteq [1:p]\mapsto f(S \mid [p+1:n])$, which is also a submodular function:
\begin{align}\label{eqn:befrthm1condhan}
     f([1:p]\mid [p+1:n]) \!\leq\! \frac{1}{\binom{p-1}{k-1} } \! \sum\limits_{\substack{S \subseteq [1:p]: \\ |S| = k}}\! \!f(S\mid [p+1:n]).
\end{align}
The key contribution in the theorem below is the second inequality, which shows that the resulting upper bound on $f([1:n])$ obtained from \eqref{eqn:befrthm1condhan} is tighter than the one in \eqref{eq: HanIneq}, and the establishment of its corresponding equality conditions.
\begin{theorem}\label{StrongHanInequalitySubmodular}
    Let $f$ be any submodular function with $f(\emptyset) = 0$. Then, for $k,p \in [1:n]$ such that $k \leq p$,
    \begin{align}
        f([1:n]) &\leq \frac{1}{\binom{p-1}{k-1}} \sum\limits_{\substack{S \subseteq [1:p]: \\ |S| = k}} f(S \mid [p+1:n]) + f([p+1:n]) \label{eq: StrongHan}\\
        & \leq \frac{1}{\binom{n-1}{k-1} }  \sum\limits_{\substack{S \subseteq [1:n]: \\ |S| = k}} f(S)\label{eq: StrongHan1}.
    \end{align}
    For $k<p$, equality in \eqref{eq: StrongHan} holds if and only if $f(S \mid [p+1:n]) = \sum\limits_{i \in S} f(\{i\} \mid [p+1:n]),~\forall S \subseteq [1:p]$. If $k=p$, equality holds trivially in \eqref{eq: StrongHan}.
    
    \noindent For $p<n$, equality in \eqref{eq: StrongHan1} holds if and only if 
    \begin{enumerate}[label=(\roman*), leftmargin=*, align=left, widest=iii, nosep]
        \item $f(S)=\sum_{i\in S}f(\{i\})$, for all $S\subseteq [1:p]$ with $|S|\leq k$,
        \item $f(S \cup [p+1:n])=f(S)+f([p+1:n])$, for all $S\subseteq [1:p]$ with $|S|\leq k$, and
        \item $f(S)=\sum_{i\in S}f(\{i\})$, for all $S\subseteq [p+1:n]$.
    \end{enumerate}
    For $p=n$, equality holds trivially in \eqref{eq: StrongHan1}.
\end{theorem}

\begin{remark}
    Theorem~\ref{StrongHanInequalitySubmodular} establishes that the upper bound on $f([1:n])$ in \eqref{eq: StrongHan} is tighter than the one in the classical form of Han's inequality for submodular functions in \eqref{eq: HanIneq}. As discussed around \eqref{eqn:befrthm1condhan}, this improvement is achieved through a conditional version of Han's inequality. This generalizes the mechanism by which an upper bound on $h(X_{[1:n]})$ is obtained through conditional subadditivity in \eqref{eqn:condsubadd}, which is tighter than the bound obtained via standard subadditivity in \eqref{eqn:subadditivity}. 
\end{remark}
\begin{remark}
    From Theorem~\ref{StrongHanInequalitySubmodular}, we can infer that the upper bound on $f([1:n])$ in \eqref{eq: StrongHan} is strictly tighter than the one in Han's inequality in \eqref{eq: HanIneq} if and only if at least one of the conditions in \textit{(i)-(iii)} is violated.
\end{remark}

A detailed proof of Theorem~\ref{StrongHanInequalitySubmodular} is provided in
\if \extended 1%
Appendix~\ref{appendix: a}.
\fi
\if \extended 0%
\cite[Appendix~A]{JakharKCP26}.
\fi 
% Theorem~\ref{StrongHanInequalitySubmodular} is applicable to any submodular function, we note some examples below: 
% \begin{itemize}[leftmargin=*]
%     \item {Entropy and differential entropy~\cite{Fujishige05}}: For $S\subseteq [1:n]$, $f(S) = H(X_S)$ or $f(S)=h(X_S)$.
%     \item {Graph-cut~\cite{Fujishige05,Bach13}}: For a graph $G = (V,E)$ and weight function $w: E \rightarrow \mathbb{R}$, $f(S) = \sum\limits_{\substack{\{u,v\} \in E: u \in S, v \notin S}} w(\{u,v\})$, for $S\subseteq V$.
%     \item {Set Cover}~\cite{Bach13,IyerKBA22}: For a universe $U$ with weights $w_u \geq 0$ and subsets $U_i \subseteq U$ for $i \in [1:n]$, $f(S) = \sum_{u \in \bigcup_{i \in S} U_i} w_u$, for $S \subseteq [1:n]$.
%      \item Matroid rank~\cite{Fujishige05,Bach13}: For a matroid $\mathcal{M} = (E, \mathcal{I})$, where $E$ is the ground set and $\mathcal{I}$ is the collection of independent sets, the rank function is $f(S) = \max_{I \in \mathcal{I}: I \subseteq S} |I|$, for $S\subseteq E$.
%     \item Facility location~\cite{Frieze74,IyerKBA22}: $f(S) =\sum_{i \in [1:n]} \max_{j \in S} s_{ij}$, for $S\subseteq [1:n]$, where $s_{ij}$ denotes similarity value between the items $i,j\in[1:n]$.
% \end{itemize}
Theorem~\ref{StrongHanInequalitySubmodular} is applicable to any submodular function, for example, entropy and differential entropy~\cite{Fujishige05}, graph-cut~\cite{Fujishige05,Bach13}, set cover~\cite{Bach13,IyerKBA22}, matroid rank~\cite{Fujishige05,Bach13}, facility location~\cite{Frieze74,IyerKBA22}, etc. Next, we specialize Theorem~\ref{StrongHanInequalitySubmodular} to differential entropy. This will be used in Section~\ref{DeterminantSection} to prove a stronger inequality than Sz\'asz's inequality for matrix determinants. A detailed proof of Proposition~\ref{StrongHanInequality} is provided in
\if \extended 1%
Appendix~\ref{appendix: b}.
\fi
\if \extended 0%
\cite[Appendix~B]{JakharKCP26}.
\fi
\begin{proposition}\label{StrongHanInequality}
    For jointly distributed continuous random variables $X_1, \ldots, X_n$, and $k,p \in [1:n]$ such that $k \leq p$,
    \begin{align}
        h(X_{[1:n]}) &\leq \frac{1}{\binom{p-1}{k-1}} \sum\limits_{\substack{S \subseteq [1:p]: \\ |S| = k}} h(X_{S}| X_{[p+1:n]}) + h(X_{[p+1:n]}) \label{eq: StrongHanEntropy}\\
        & \leq \frac{1}{\binom{n-1}{k-1} }  \sum\limits_{\substack{S \subseteq [1:n]: \\ |S| = k}} h(X_S). \label{eq: StrongHanEntropy1}
    \end{align}
    For $k<p$, equality in \eqref{eq: StrongHanEntropy} holds if and only if the random variables $X_1,\dots,X_p$ are conditionally independent given $X_{[p+1:n]}$. If $k=p$, equality holds trivially in \eqref{eq: StrongHanEntropy}.
    
     \noindent For $p<n$, equality in \eqref{eq: StrongHanEntropy1} holds if and only if
    \begin{enumerate}[label=(\roman*), leftmargin=*, align=left, widest=iii, nosep]
        \item any $k$ random variables taken from $\{X_1, \ldots, X_p\}$ are mutually independent,
        \item any $k$ random variables taken from $\{X_1, \ldots, X_p\}$ are independent of $X_{[p+1:n]}$, and
        \item $X_{p+1},\dots,X_n$ are mutually independent.
    \end{enumerate}
    For $p=n$, equality holds trivially in \eqref{eq: StrongHanEntropy1}.
\end{proposition}

\begin{remark}\label{HanStrongerThanSubadd}
   For $2 \leq k \leq n$, it follows from the monotonicity property of Han's inequality that the upper bound on $h(X_{[1:n]})$ in \eqref{eq: StrongHanEntropy} is tighter than the bound for the $k=1$ case, which corresponds to the conditional subadditivity of entropy in \eqref{eqn:condsubadd}. See
    \if \extended 1%
    Appendix~\ref{appendix: c}
    \fi
    \if \extended 0%
    \cite[Appendix~C]{JakharKCP26}
    \fi
    for details.
\end{remark}

The conditioning mechanism used to strengthen Han's inequality for $k$-sized subset families, yielding Theorem~\ref{StrongHanInequalitySubmodular}, can be applied to partitions of $[1:n]$ as well, resulting in a tighter bound than standard partition subadditivity in \eqref{eq: PartitionSubadditivity}. 

\begin{theorem}\label{ConditionalSubadditivitySubmodular}
    Let $f$ be any submodular function with $f(\emptyset) = 0$, and let $\mathcal{P}$ be any partition of $[1:n]$. Then, for any $p \in [1:n]$, and the induced partition $\mathcal{P}'$ of $[1:p]$ such that $\mathcal{P}' = \{ S \cap [1:p]: S \in \mathcal{P},\ S \cap [1:p]\neq \emptyset \}$, 
    \begin{align}
        f([1:n]) &\leq \sum\limits_{S \in \mathcal{P}'} f(S \mid [p+1:n]) + f([p+1:n]) \label{eq: conditionalSub1} \\
        & \leq \sum\limits_{S \in \mathcal{P}} f(S). \label{eq: conditionalSub2}
    \end{align}
    Equality in \eqref{eq: conditionalSub1} holds if and only if $f([1:p] \mid [p+1:n]) = \sum\limits_{S\in \mathcal{P}'} f(S \mid [p+1:n])$. 
    
    \noindent Equality in \eqref{eq: conditionalSub2} holds if and only if:
    \begin{enumerate}[label=(\roman*), leftmargin=*, align=left, widest=ii, nosep]
        \item $f(S) = f(S \cap [p+1:n]) + f(S \cap [1:p] \mid [p+1:n])$, for all $S \in \mathcal{P}$, and
        \item $f([p+1:n])=\sum_{S\in\mathcal{P}}f(S\cap [p+1:n])$.
    \end{enumerate}
\end{theorem}

A detailed proof of Theorem~\ref{ConditionalSubadditivitySubmodular} is provided in
\if \extended 1%
Appendix~\ref{appendix: d}.
\fi
\if \extended 0%
\cite[Appendix~D]{JakharKCP26}.
\fi
We specialize Theorem~\ref{ConditionalSubadditivitySubmodular} to differential entropy, which will be used in Section~\ref{DeterminantSection} to prove a stronger inequality than Fischer's inequality for matrix determinants. A detailed proof of Proposition~\ref{ConditionalSubadditivity} is provided in
\if \extended 1%
Appendix~\ref{appendix: e}.
\fi
\if \extended 0%
\cite[Appendix~E]{JakharKCP26}.
\fi

\begin{proposition}\label{ConditionalSubadditivity}
    Let $X_1, \ldots, X_n$ be jointly distributed continuous random variables. Then, for any $p \in [1:n]$, a partition $\mathcal{P}$ of $[1:n]$ and the induced partition $\mathcal{P}'$ of $[1:p]$ such that $\mathcal{P}' = \{ S \cap [1:p]: S \in \mathcal{P}, \ S \cap [1:p]\neq \emptyset\}$, 
    \begin{align}
        h(X_{[1:n]}) &\leq \sum\limits_{S \in \mathcal{P}'} h(X_S|X_{[p+1:n]}) + h(X_{[p+1:n]}) \label{eq: conditionalSubEntropy1} \\
        & \leq \sum\limits_{S \in \mathcal{P}} h(X_S). \label{eq: conditionalSubEntropy2}
    \end{align}
    Equality in \eqref{eq: conditionalSubEntropy1} holds if and only if $X_{S}$, for $S\in\mathcal{P}'$ are conditionally independent given $X_{[p+1:n]}$. 
    
    \noindent Equality in \eqref{eq: conditionalSubEntropy2} holds if and only if:
    \begin{enumerate}[label=(\roman*), leftmargin=*, align=left, widest=ii, nosep]
        \item $X_{S \cap [1:p]}$ and $X_{S^{\mathrm{c}}\cap[p+1:n]}$ are conditionally independent given $X_{S\cap[p+1:n]}$, for all $S\in\mathcal{P}$, and
        \item $X_{S\cap [p+1:n]}$, for $S\in\mathcal{P}$ are independent.
    \end{enumerate}
    
\end{proposition}

\section{Strengthening Sz\'asz's and Fischer's Inequalities for Matrix Determinants}\label{DeterminantSection}
In this section, we apply the differential entropy bounds in Propositions~\ref{StrongHanInequality} and \ref{ConditionalSubadditivity} to a zero-mean Gaussian random vector. This yields strengthened versions of classical Sz\'asz's and Fischer's determinantal inequalities previously stated in Theorem~\ref{thm:matrixineq}.

\begin{theorem}\label{StrongSzasz}
    Let $K$ be a positive definite $n \times n$ matrix. Then, for any $k,p \in [1:n]$ such that $k \leq p$, and $P = [1:p]$,
    \begin{align}
        |K| & \leq |K(P^{\mathrm{c}})|\bigg( \prod\limits_{S \subseteq P:|S|=k} \frac{|K(S \cup P^{\mathrm{c}})|}{|K(P^{\mathrm{c}})|}\bigg)^{\frac{1}{\binom{p-1}{k-1}}} \label{eq: SS_1}\\
        & \leq \bigg(\prod\limits_{S \subseteq [1:n]:|S|=k} |K(S)|\bigg)^{\frac{1}{\binom{n-1}{k-1}}}. \label{eq: SS_2}
    \end{align}
    For $k < p$, equality in \eqref{eq: SS_1} holds if and only if the Schur complement of $K(P^{\mathrm{c}})$ in $K$, i.e., $K(P) - K(P,P^{\mathrm{c}}) {K(P^{\mathrm{c}})}^{-1} K(P^{\mathrm{c}},P)$ is diagonal. 
    
    \noindent For $k = 1$ and $p<n$, equality in \eqref{eq: SS_2} holds if and only if $K(P,P^\mathrm{c})=0$, and $K(P^{\mathrm{c}})$ is diagonal. For $2 \leq k \leq p<n$, equality in \eqref{eq: SS_2} holds if and only if $K$ is diagonal. For $p=n$, equality holds trivially in \eqref{eq: SS_2}.
\end{theorem}

\begin{remark}\label{remark:szaszsrtong}
    Theorem~\ref{StrongSzasz} establishes that the upper bound on $|K|$ in \eqref{eq: SS_1} is tighter than the one provided by Sz\'{a}sz's inequality in \eqref{eq: Szasz}. Moreover, from the equality conditions of \eqref{eq: SS_2} for $2\leq k \leq p<n$, it follows that the bound in \eqref{eq: SS_1} is strictly tighter than the one in Sz\'{a}sz's inequality whenever $K$ is a non-diagonal matrix. Notably, equality in \eqref{eq: SS_2} implies that equality in \eqref{eq: SS_1} holds---a phenomenon that does not hold in general for Proposition~\ref{StrongHanInequality} or Theorem~\ref{StrongHanInequalitySubmodular}. This occurs because the equality conditions \textit{(i)-(iii)} in Proposition~\ref{StrongHanInequality} are equivalent to mutual independence of all the random variables $X_1,\dots,X_n$ in the Gaussian setting.
\end{remark}

\begin{remark}\label{remark:StrngSzaszFan}
    Theorem~\ref{StrongSzasz} recovers Ky Fan's inequality in \eqref{eq: KyFan} when $k=1$. Furthermore, in view of the discussion in Remark~\ref{HanStrongerThanSubadd}, the bound corresponding to $2 \leq k \leq p$ in \eqref{eq: SS_1} is tighter than the bound obtained for $k=1$. A consequence of the equality conditions of \eqref{eq: SS_2} for $k=1$ is that Ky Fan's inequality is strictly stronger than Hadamard's inequality whenever either $K(P,P^{\mathrm{c}}) \neq 0$ or $K(P^{\mathrm{c}})$ is a non-diagonal matrix.
\end{remark}

\begin{remark}\label{LinAlgApproach}
    We note that Ky Fan's inequality in \eqref{eq: KyFan} was originally proved in \cite[Theorem~1]{Fan55} by applying Hadamard's inequality to the Schur complement of $K([p+1:n])$ in $K$. Generalizing Ky Fan's approach, in 
    \if \extended 1%
    Appendix~\ref{appendix: g},
    \fi
    \if \extended 0%
    \cite[Appendix~G]{JakharKCP26},
    \fi we provide an alternative proof of \eqref{eq: SS_1} by applying Sz\'asz's inequality to the same Schur complement. However, we do not know of a direct proof of the comparison in \eqref{eq: SS_2} using this algebraic approach.
\end{remark}

A detailed proof of Theorem~\ref{StrongSzasz} is provided in
\if \extended 1%
Appendix~\ref{appendix: f}.
\fi
\if \extended 0%
\cite[Appendix~F]{JakharKCP26}.
\fi
The following example illustrates that the bound in \eqref{eq: SS_1} is tighter than those obtained from both Sz\'asz's and Ky Fan's inequalities.

\begin{example}\label{eg:1}
    Consider the positive definite matrix 
    \begin{equation}\label{eqn:matrix}
        A = \begin{pmatrix}
        2 & 1 & 1 & 1\\
        1 & 3 & 1 & 1\\
        1 & 1 & 4 & 1\\
        1 & 1 & 1 & 5
        \end{pmatrix}.
    \end{equation}
    Applying Sz\'asz's inequality in \eqref{eq: Szasz} with $k=2$ gives
    \begin{align}
        |A| &\leq \bigg( \prod_{S \subseteq [1:4]: |S| = 2} |A(S)| \bigg)^{1/3} = 97.32.
    \end{align}
    Applying Ky Fan's inequality in \eqref{eq: KyFan} with $p=3$ gives
    \begin{align}
        |A| &\leq \frac{|A(\{1,4\})| |A(\{2,4\})| |A(\{3,4\})|}{|A(\{4\})|^2} = 95.76.
    \end{align}
    Now, by applying the inequality in \eqref{eq: SS_1} with $k=2$ and $p=3$, we obtain
    \begin{align}
        |A| &\leq \bigg(\frac{|A(\{1,2,4\})| |A(\{2,3,4\})| |A(\{1,3,4\})|}{|A(\{4\})|}\bigg)^{1/2} \nonumber \\
        &= 82.58.
    \end{align}
    This bound is strictly tighter than both of the aforementioned bounds, and is closer to the actual determinant value of $|A|=74$.
\end{example}

Motivated by an analogous result of Ky Fan~\cite[Theorem~4]{Fan55}, we generalize Theorem~\ref{StrongSzasz} to establish an inequality concerning the eigenvalues of positive definite matrices.

\begin{theorem}\label{EigenvaluesStrongKyFan}
    Let $K$ be a positive definite $n \times n$ matrix and let $\lambda_1 \leq \lambda_2 \leq \ldots \leq \lambda_n$ be its eigenvalues arranged in increasing order. Let $h,p,k \in [1:n]$ and $\ell\in \{0\}\cup[1:n]$ be such that $1 \leq k \leq h \leq p \leq n$, $0 \leq \ell \leq n-p$, $P = [1:p]$ and $Q_\ell = [p+1:p+\ell]$\footnote{We adopt the convention that $Q_\ell = \emptyset$ when $\ell = 0$, and recall that $|K(\emptyset)| = 1$.}. Then
    \begin{align}
        \prod\limits_{i=1}^{h+\ell} \lambda_i \leq |K(Q_\ell)| \bigg( \prod\limits_{S \subseteq [1:h]: |S|=k} \frac{|K(S \cup P^{\mathrm{c}})|}{|K(P^{\mathrm{c}})|}\bigg)^{\frac{1}{\binom{h-1}{k-1}}}\label{eqn:eigen}.
    \end{align}
\end{theorem}

A detailed proof of Theorem~\ref{EigenvaluesStrongKyFan} is provided in
\if \extended 1%
Appendix~\ref{appendix: h}.
\fi
\if \extended 0%
\cite[Appendix~H]{JakharKCP26}.
\fi

\begin{remark}
    Theorem~\ref{EigenvaluesStrongKyFan} recovers \cite[Theorem~4]{Fan55} when $k=1$. Furthermore, in view of the discussion in Remark~\ref{remark:StrngSzaszFan}, the bound corresponding to $2 \leq k \leq h$ in \eqref{eqn:eigen} is tighter than the bound obtained for $k=1$. We illustrate this strengthening in the following example.
\end{remark}

\begin{example}\label{eg:2}
    Consider the positive definite matrix $A$ in \eqref{eqn:matrix} from Example~\ref{eg:1}. 

    Applying the eigenvalue inequality of Ky Fan~\cite[Theorem~4]{Fan55} for the product of $h=3$ smallest eigenvalues of $A$ with $p=3$ gives 
    \begin{align}\label{eqn:exmpleeigenv}
        \lambda_1\lambda_2\lambda_3 \leq \frac{|A(\{1,4\})| |A(\{2,4\})| |A(\{3,4\})|}{|A(\{4\})|^3}= 19.152
    \end{align}
    Now, by applying the inequality in \eqref{eqn:eigen} with $h=3$, $p=3$, $\ell=0$, $k=2$, we obtain
    \begin{align}
        \lambda_1\lambda_2\lambda_3 &\leq 1 \times \bigg(\frac{|A(\{1,2,4\})| |A(\{2,3,4\})| |A(\{1,3,4\})|}{|A(\{4\})|^3}\bigg)^{1/2} \nonumber \\
        &= 16.516.
    \end{align}
    This bound is strictly tighter than the one in \eqref{eqn:exmpleeigenv} and is closer to the actual value, $\lambda_1\lambda_2\lambda_3 \approx 10.876$.
\end{example}

\begin{theorem}\label{StrongFischer}
    Let $K$ be a positive definite $n \times n$ matrix. Then, for any $p \in [1:n], P = [1:p]$, a partition $\mathcal{P}$ of $[1:n]$, and the induced partition $\mathcal{P}'$ of $P$ such that $\mathcal{P}' = \{ S \cap P: S \in \mathcal{P}, \ S \cap [1:p]\neq \emptyset\}$,
    \begin{align}
        |K| & \leq |K(P^{\mathrm{c}})| \prod\limits_{S \in \mathcal{P}'} \frac{|K(S \cup P^{\mathrm{c}})|}{|K(P^{\mathrm{c}})|} \label{eq: SF_1} \\
        & \leq \prod\limits_{S \in \mathcal{P}} |K(S)|. \label{eq: SF_2}
    \end{align}
    Equality in \eqref{eq: SF_1} holds if and only if $M(S, S') = 0,$ for all distinct $S, S' \in \mathcal{P}'$, where $M$ is the Schur complement of $K(P^{\mathrm{c}})$ in $K$, i.e., $M = K(P) - K(P,P^{\mathrm{c}}) K(P^{\mathrm{c}})^{-1} K(P^{\mathrm{c}},P)$.
    
    \noindent Equality in \eqref{eq: SF_2} holds if and only if
    \begin{enumerate}[label=(\roman*), leftmargin=*, align=left, widest=ii, nosep]
        \item for all $S \in \mathcal{P}$, let
    $A = S \cap P$, $B = S \cap P^{\mathrm{c}}$, and
    $C = S^{\mathrm{c}} \cap P^{\mathrm{c}}$. Then
    \[
    K(A,C) =
    \begin{cases}
    K(A,B)K(B)^{-1}K(B,C), & B \neq \emptyset\\
    0, & B = \emptyset
    \end{cases},
    \]
       
        \item $K(S \cap P^{\mathrm{c}}, S' \cap P^{\mathrm{c}}) = K(S' \cap P^{\mathrm{c}},S \cap P^{\mathrm{c}}) = 0,$ for all distinct $S,S' \in \mathcal{P}$.
    \end{enumerate}
\end{theorem}
A detailed proof of Theorem~\ref{StrongFischer} is provided in
\if \extended 1%
Appendix~\ref{appendix: i}.
\fi
\if \extended 0%
\cite[Appendix~I]{JakharKCP26}.
\fi

\begin{example}\label{eg:4}
    Consider the positive definite matrix $A$ in \eqref{eqn:matrix} from Example~\ref{eg:1}. Applying Fischer's inequality with $\mathcal{P} = \{\{1,3\},\{2,4\}\}$ gives 
    \begin{align}
        |A| \leq |A(\{1,3\})||A(\{2,4\})| = 5 \times 19 = 95.
    \end{align}
    Now, by applying the inequality in \eqref{eq: SF_1} with $p=2$ for the same $\mathcal{P}$ gives $|A| \leq   81.58$.
     % \begin{align}
     %      |A| \leq \frac{|A(\{1,3,4\})||A(\{2,3,4\})|}{|A(\{3,4\})|} = 81.58.
     % \end{align}
   
\end{example}

% \section*{Acknowledgment}

% We are indebted to Michael Shell for maintaining and improving
% \texttt{IEEEtran.cls}. 

%%%%%%
%% To balance the columns at the last page of the paper use this
%% command:
%%
%\enlargethispage{-1.2cm} 
%%
%% If the balancing should occur in the middle of the references, use
%% the following trigger:
%%
\if \extended 0%
\IEEEtriggeratref{12}
\fi

\if \extended 1%
\IEEEtriggeratref{13}
\fi
%%
%% which triggers a \newpage (i.e., new column) just before the given
%% reference number. Note that you need to adapt this if you modify
%% the paper.  The "triggered" command can be changed if desired:
%%
%\IEEEtriggercmd{\enlargethispage{-20cm}}
%%
%%%%%%

% \clearpage
% \pagebreak

%%%%%%
%% References:
%% We recommend the usage of BibTeX:
%%
\bibliographystyle{IEEEtran}
\bibliography{bibliofile}

\begin{thebibliography}{10}
\providecommand{\url}[1]{#1}
\csname url@samestyle\endcsname
\providecommand{\newblock}{\relax}
\providecommand{\bibinfo}[2]{#2}
\providecommand{\BIBentrySTDinterwordspacing}{\spaceskip=0pt\relax}
\providecommand{\BIBentryALTinterwordstretchfactor}{4}
\providecommand{\BIBentryALTinterwordspacing}{\spaceskip=\fontdimen2\font plus
\BIBentryALTinterwordstretchfactor\fontdimen3\font minus
  \fontdimen4\font\relax}
\providecommand{\BIBforeignlanguage}[2]{{%
\expandafter\ifx\csname l@#1\endcsname\relax
\typeout{** WARNING: IEEEtran.bst: No hyphenation pattern has been}%
\typeout{** loaded for the language `#1'. Using the pattern for}%
\typeout{** the default language instead.}%
\else
\language=\csname l@#1\endcsname
\fi
#2}}
\providecommand{\BIBdecl}{\relax}
\BIBdecl

\bibitem{Fujishige05}
S.~Fujishige, \emph{Submodular functions and optimization}.\hskip 1em plus
  0.5em minus 0.4em\relax Elsevier, 2005.

\bibitem{krause2007near}
A.~Krause and C.~Guestrin, ``Near-optimal observation selection using
  submodular functions,'' in \emph{AAAI}, vol.~7, 2007, pp. 1650--1654.

\bibitem{ConfortiC84}
M.~Conforti and G.~Cornuéjols, ``Submodular set functions, matroids and the
  greedy algorithm: Tight worst-case bounds and some generalizations of the
  {R}ado-{E}dmonds theorem,'' \emph{Discrete Applied Mathematics}, vol.~7,
  no.~3, pp. 251--274, 1984.

\bibitem{lehmann2001combinatorial}
B.~Lehmann, D.~Lehmann, and N.~Nisan, ``Combinatorial auctions with decreasing
  marginal utilities,'' in \emph{Proceedings of the 3rd ACM conference on
  Electronic Commerce}, 2001, pp. 18--28.

\bibitem{kempe2003maximizing}
D.~Kempe, J.~Kleinberg, and {\'E}.~Tardos, ``Maximizing the spread of influence
  through a social network,'' in \emph{Proceedings of the ninth ACM SIGKDD
  international conference on Knowledge discovery and data mining}, 2003, pp.
  137--146.

\bibitem{VanEnterFS93}
A.~C. Van~Enter, R.~Fern{\'a}ndez, and A.~D. Sokal, ``Regularity properties and
  pathologies of position-space renormalization-group transformations: Scope
  and limitations of {G}ibbsian theory,'' \emph{Journal of Statistical
  Physics}, vol.~72, pp. 879--1167, 1993.

\bibitem{Bach13}
F.~Bach, \emph{Learning with Submodular Functions: A Convex Optimization
  Perspective}.\hskip 1em plus 0.5em minus 0.4em\relax Now Publishers Inc.,
  2013.

\bibitem{IyerKBA22}
R.~Iyer, N.~Khargonkar, J.~Bilmes, and H.~Asnani, ``Generalized submodular
  information measures: Theoretical properties, examples, optimization
  algorithms, and applications,'' \emph{IEEE Transactions on Information
  Theory}, vol.~68, no.~2, pp. 752--781, 2022.

\bibitem{Han78}
T.~S. Han, ``Nonnegative entropy measures of multivariate symmetric
  correlations,'' \emph{Information and Control}, vol.~36, no.~2, pp. 133--156,
  1978.

\bibitem{BoucheronLM13}
S.~Boucheron, G.~Lugosi, and P.~Massart, \emph{Concentration Inequalities: A
  Nonasymptotic Theory of Independence}.\hskip 1em plus 0.5em minus 0.4em\relax
  Oxford University Press, 02 2013.

\bibitem{Sason22}
\BIBentryALTinterwordspacing
I.~Sason, ``Information inequalities via submodularity and a problem in
  extremal graph theory,'' \emph{Entropy}, vol.~24, no.~5, 2022. [Online].
  Available: \url{https://www.mdpi.com/1099-4300/24/5/597}
\BIBentrySTDinterwordspacing

\bibitem{BarYossefJKS02}
Z.~Bar-Yossef, T.~Jayram, R.~Kumar, and D.~Sivakumar, ``Information theory
  methods in communication complexity,'' in \emph{Proceedings 17th IEEE Annual
  Conference on Computational Complexity}, 2002, pp. 93--102.

\bibitem{SunJ18}
H.~Sun and S.~A. Jafar, ``The capacity of robust private information retrieval
  with colluding databases,'' \emph{IEEE Transactions on Information Theory},
  vol.~64, no.~4, pp. 2361--2370, 2018.

\bibitem{CarlenC09}
E.~A. Carlen and D.~Cordero-Erausquin, ``Subadditivity of the entropy and its
  relation to {B}rascamp--{L}ieb type inequalities,'' \emph{Geometric and
  Functional Analysis}, vol.~19, no.~2, pp. 373--405, Sep 2009.

\bibitem{Pippenger99}
N.~Pippenger, ``An information-theoretic method in combinatorial theory,''
  \emph{Journal of Combinatorial Theory, Series A}, vol.~23, no.~1, pp.
  99--104, 1977.

\bibitem{MadimanT10}
M.~Madiman and P.~Tetali, ``Information inequalities for joint distributions,
  with interpretations and applications,'' \emph{IEEE Transactions on
  Information Theory}, vol.~56, no.~6, pp. 2699--2713, 2010.

\bibitem{cover1983information}
T.~Cover and A.~Gamal, ``An information-theoretic proof of {H}adamard's
  inequality (corresp.),'' \emph{IEEE Transactions on Information Theory},
  vol.~29, no.~6, pp. 930--931, 1983.

\bibitem{DemboCT91}
A.~Dembo, T.~Cover, and J.~Thomas, ``Information theoretic inequalities,''
  \emph{IEEE Transactions on Information Theory}, vol.~37, no.~6, pp.
  1501--1518, 1991.

\bibitem{HornJ12}
R.~A. Horn and C.~R. Johnson, \emph{Matrix analysis}.\hskip 1em plus 0.5em
  minus 0.4em\relax Cambridge university press, 2012.

\bibitem{Fan55}
K.~Fan, ``Some inequalities concerning positive-definite {H}ermitian
  matrices,'' \emph{Mathematical Proceedings of the Cambridge Philosophical
  Society}, vol.~51, no.~3, p. 414–421, 1955.

\bibitem{Fujishige78}
S.~Fujishige, ``Polymatroidal dependence structure of a set of random
  variables,'' \emph{Information and Control}, vol.~39, no.~1, pp. 55--72,
  1978.

\bibitem{Frieze74}
\BIBentryALTinterwordspacing
A.~M. Frieze, ``A cost function property for plant location problems,''
  \emph{Mathematical Programming}, vol.~7, no.~1, pp. 245--248, Dec 1974.
  [Online]. Available: \url{https://doi.org/10.1007/BF01585521}
\BIBentrySTDinterwordspacing

\bibitem{JakharKCP25}
G.~Jakhar, G.~R. Kurri, S.~Chillara, and V.~M. Prabhakaran, ``Fractional
  subadditivity of submodular functions: Equality conditions and their
  applications,'' in \emph{IEEE International Symposium on Information Theory
  (ISIT)}, 2025, pp. 1--6.

\bibitem{Fan50}
K.~Fan, ``On a theorem of {W}eyl concerning eigenvalues of linear
  transformations. ii,'' \emph{Proceedings of the National Academy of Sciences
  of the United States of America}, vol.~36, no.~1, pp. 31--35, 1950.

\end{thebibliography}
\clearpage

%%
%% where we here have assumed the existence of the files
%% definitions.bib and bibliofile.bib.
%% BibTeX documentation can be obtained at:
%% http://www.ctan.org/tex-archive/biblio/bibtex/contrib/doc/
%%%%%%

% Moreover, example code for an appendix (or appendices) can also be
% found in the source file (they are commented out).

%%%%%%
%% Appendix:
%% If needed a single appendix is created by
%%
%\appendix
%%
%% If several appendices are needed, then the command
%%

\if \extended 1

\appendices
%%
%% in combination with further \section commands can be used.
%%%%%%

\section{Proof of Theorem~\ref{StrongHanInequalitySubmodular}}\label{appendix: a}
Let $U \subseteq [1:n]$ such that $|U| = k$. Note that $U$ can be uniquely partitioned into $U = S \cup T$, where $S \subseteq [1:p],T \subseteq [p+1:n]$ such that $|S|=m,|T|=k-m$, for some integer $m \in [0:k]$. By chain rule~\eqref{eqref: chain_rule}, we get
\begin{align}
    f(U) = f(S \cup T) & = f(S \mid T) + f(T)\\
    &\geq f(S \mid [p+1:n]) + f(T) \label{eq: shis_2},
\end{align}
where \eqref{eq: shis_2} follows since $T \subseteq [p+1:n]$ implies $f(S \mid [p+1:n]) \leq f(S \mid T)$ by \eqref{eqref: conditioning}. Summing \eqref{eq: shis_2} over all $\binom{n}{k}$ subsets $U \subseteq [1:n]$ of size $k$ yields
\begin{align}
    &\sum_{\substack{U \subseteq [1:n] \\ |U|=k}} f(U) \nonumber \\
    &\geq \sum_{m=0}^k \sum_{\substack{S \subseteq [1:p] \\ |S|=m}} \sum_{\substack{T \subseteq [p+1:n] \\ |T|=k-m}} \Big[ f(S \mid [p+1:n]) + f(T) \Big] \label{eq: shis_15}\\
    &= \sum_{m=0}^k \sum_{\substack{S \subseteq [1:p] \\ |S|=m}} \sum_{\substack{T \subseteq [p+1:n] \\ |T|=k-m}} f(S \mid [p+1:n]) \nonumber \\
    & \quad\quad + \sum_{m=0}^k \sum_{\substack{S \subseteq [1:p] \\ |S|=m}} \sum_{\substack{T \subseteq [p+1:n] \\ |T|=k-m}} f(T). \label{eq: shis_3}
\end{align}
We now lower bound both the triple summations on the RHS of \eqref{eq: shis_3}. The first triple summation can be lower bounded as follows.
\begin{align}
    & \sum_{m=0}^k \sum_{\substack{S \subseteq [1:p] \\ |S|=m}} \sum_{\substack{T \subseteq [p+1:n] \\ |T|=k-m}} f(S \mid [p+1:n]) \nonumber \\
    & = \sum\limits_{m = 0}^k \sum_{\substack{S \subseteq [1:p] \\ |S|=m}} f(S \mid [p+1:n]) \left(\sum_{\substack{T \subseteq [p+1:n] \\ |T|=k-m}}1\right) \\
    & = \sum\limits_{m = 0}^k \binom{n-p}{k-m} \sum_{\substack{S \subseteq [1:p] \\ |S|=m}} f(S \mid [p+1:n]) \label{eq: shis_4}\\
    & = \sum\limits_{m = 1}^k \binom{n-p}{k-m} \sum_{\substack{S \subseteq [1:p] \\ |S|=m}} f(S \mid [p+1:n]) \label{eq: shis_5}\\
    & \geq \sum\limits_{m = 1}^k \binom{n-p}{k-m}\binom{p-1}{m-1} \frac{1}{\binom{p-1}{k-1}} \sum_{\substack{S \subseteq [1:p] \\ |S|=k}} f(S \mid [p+1:n]) \label{eq: shis_6}\\
    & = \frac{1}{\binom{p-1}{k-1}} \sum_{\substack{S \subseteq [1:p] \\ |S|=k}} f(S \mid [p+1:n]) \left( \sum\limits_{m = 1}^k \binom{n-p}{k-m}\binom{p-1}{m-1} \right) \\
    & = \frac{\binom{n-1}{k-1}}{\binom{p-1}{k-1}} \sum_{\substack{S \subseteq [1:p] \\ |S|=k}} f(S \mid [p+1:n]) \label{eq: shis_7},
\end{align}
where \eqref{eq: shis_4} follows since there are $\tbinom{n-p}{k-m}$ subsets $T \subseteq [p+1:n$ such that $|T| = k-m$, \eqref{eq: shis_5} follows because $f(\emptyset \mid [p+1:n]) = f([p+1:n]) - f([p+1:n]) = 0$, \eqref{eq: shis_6} follows from the monotonically decreasing nature of the sequence of average of any submodular set function $g$ values over all subsets of same size~\cite{Han78,Sason22}\footnote{\eqref{eq: shis_8} is a special case of $\alpha = 1$ in \cite[Corollary~2]{Sason22} whereas the exact analogue of this sequence result was initially proved for $g(S) = h(X_S),~S \subseteq [1:n]$ by Han in \cite{Han78}.}, i.e.,
\begin{align}\label{eq: shis_8}
    \sum\limits_{\substack{S \subseteq [1:p]:\\|S| = k}} \frac{g(S)}{\binom{p}{k}k} \leq \sum\limits_{\substack{S \subseteq [1:p]:\\|S| = m}} \frac{g(S)}{\binom{p}{m}m},~m < k,
\end{align}  
applied, in particular, to the contracted submodular set function $g: S\subseteq [1:p] \mapsto f(S \mid [p+1:n])$ and the fact that $\binom{p}{\ell} = \frac{p}{\ell} \binom{p-1}{\ell-1},$ for any positive integer $\ell \leq p$, and \eqref{eq: shis_7} follows by Vandermonde's convolution identity.

Next, we lower bound the second triple summation on the RHS of \eqref{eq: shis_3} as follows.
\begin{align}
    & \sum_{m=0}^k \sum_{\substack{S \subseteq [1:p] \\ |S|=m}} \sum_{\substack{T \subseteq [p+1:n] \\ |T|=k-m}} f(T) \nonumber \\ 
    & = \sum_{m=0}^k \sum_{\substack{T \subseteq [p+1:n] \\ |T|=k-m}} f(T) \left(\sum_{\substack{S \subseteq [1:p] \\ |S|=m}} 1\right) \\
    & = \sum_{m=0}^k \binom{p}{m} \sum_{\substack{T \subseteq [p+1:n] \\ |T|=k-m}} f(T) \label{eq: shis_9}\\
    & = \sum_{j=1}^k \binom{p}{k-j} \sum_{\substack{T \subseteq [p+1:n] \\ |T|= j}} f(T) \label{eq: shis_10}\\
    & \geq \sum_{j=1}^k \binom{p}{k-j} \binom{n-p-1}{j-1} f([p+1:n]) \label{eq: shis_11}\\
    & = f([p+1:n]) \left( \sum_{j=1}^k \binom{p}{k-j} \binom{n-p-1}{j-1} \right) \\
    & = \binom{n-1}{k-1} f([p+1:n]) \label{eq: shis_12},
\end{align}
where \eqref{eq: shis_9} follows since there are $\tbinom{p}{m}$ subsets $S \subseteq [1:p$ such that $|S| = m$, \eqref{eq: shis_10} follows from substitution of variable $m$ by $k-j$ and the fact that $f(\emptyset) = 0$, and \eqref{eq: shis_11} follows from Theorem~\ref{HanInequality} applied to the contracted submodular function $g: S\subseteq [p+1:n] \mapsto f(S)$, and \eqref{eq: shis_12} follows by Vandermonde's convolution identity.

By putting together \eqref{eq: shis_3}, \eqref{eq: shis_7}, and \eqref{eq: shis_12}, we get
\begin{align}\label{eq: shis_13}
    \sum_{\substack{U \subseteq [1:n] \\ |U|=k}} f(U) &\geq \frac{\binom{n-1}{k-1}}{\binom{p-1}{k-1}} \sum_{\substack{S \subseteq [1:p] \\ |S|=k}} f(S \mid [p+1:n]) \nonumber \\
    &\quad\quad+ \binom{n-1}{k-1} f([p+1:n]).
\end{align}
Dividing both the sides by $\tbinom{n-1}{k-1}$ on both the sides of \eqref{eq: shis_13} gives
\begin{align}\label{eq: shis_14}
    \frac{1}{\binom{n-1}{k-1}}\sum_{\substack{U \subseteq [1:n] \\ |U|=k}} f(U) &\geq \frac{1}{\binom{p-1}{k-1}} \sum_{\substack{S \subseteq [1:p] \\ |S|=k}} f(S \mid [p+1:n]) \nonumber \\
    &\quad\quad+ f([p+1:n]).
\end{align}
Finally putting together \eqref{eq: shis_14} with \eqref{eq: HanIneq}, we have established
\begin{align}
    f([1:n]) &\leq \frac{1}{\binom{p-1}{k-1}} \sum\limits_{\substack{S \subseteq [1:p]: \\ |S| = k}} f(S \mid [p+1:n]) + f([p+1:n]) \\
    & \leq \frac{1}{\binom{n-1}{k-1}}  \sum\limits_{\substack{S \subseteq [1:n]: \\ |S| = k}} f(S).
\end{align}

Let us now analyze the equality conditions. So, equality in \eqref{eq: StrongHan} is equivalent to equality in
\begin{align}\label{eq: shis_27}
    f([1:p]\mid [p+1:n]) \!\leq\! \frac{1}{\binom{p-1}{k-1} } \! \sum\limits_{\substack{S \subseteq [1:p]: \\ |S| = k}}\! \!f(S\mid [p+1:n]).
\end{align}
For $k = p$, there is nothing to analyze, as both LHS and RHS are equal to $f([1:p] \mid [p+1:n])$. For $k < p$, equality in \eqref{eq: shis_27} holds if and only if the contracted submodular set function $g: S\subseteq [1:p] \mapsto f(S \mid [p+1:n])$ is modular by \cite[Theorem~4]{JakharKCP25} or in other words, $f(S \mid [p+1:n]) = \sum\limits_{i \in S} f(\{i\} \mid [p+1:n]),~\forall S \subseteq [1:p]$ for $k<p$. To analyze the necessary and sufficient conditions for equality to hold in \eqref{eq: StrongHan1}, it is clear that equality holds if and only if equality is attained in each of \eqref{eq: shis_15}, \eqref{eq: shis_6}, and \eqref{eq: shis_11}. 

It is evident that equality in \eqref{eq: shis_15} holds if and only if equality holds in \eqref{eq: shis_2} for all $S \subseteq [1:p]$ and $T \subseteq [p+1:n]$ such that $S \cup T = U$. Thus, $f(S \mid T) = f(S \mid [p+1:n]),$ for all $S \subseteq [1:p], T \subseteq [p+1:n]$ such that $S \cup T = U$. In particular, $f(S) = f(S \mid [p+1:n])$, for all $S \subseteq [1:p]$ such that $|S| = k$. Equivalently, $f(S \cup [p+1:n])=f(S)+f([p+1:n])$, for all $S\subseteq [1:p]$ such that $|S| = k$. For any $S \subseteq [1:p]$ such that $|S| = k$, and $S' \subseteq S$
\begin{align}
    0 & = f(S) + f([p+1:n]) - f(S \cup [p+1:n]) \label{eq: shis_18}\\
    & = f(S') + f(S \setminus{S'} \mid S') + f([p+1:n]) \nonumber \\
    & \quad\quad - f(S' \cup [p+1:n]) - f((S\setminus{S'}) \mid S' \cup [p+1:n]) \label{eq: shis_16}\\
    & = \underbrace{f(S') + f([p+1:n]) - f(S' \cup [p+1:n])}_{\geq 0} \nonumber \\
    & \quad\quad + \underbrace{f(S \setminus{S'} \mid S') - f((S\setminus{S'}) \mid S' \cup [p+1:n])}_{\geq 0}, \label{eq: shis_17}
\end{align}
where \eqref{eq: shis_16} follows from chain rule~\eqref{eqref: chain_rule}. Notice that both the expressions in \eqref{eq: shis_17} are non-negative because $f(S' \cup [p+1:n]) \leq f(S') + f([p+1:n])$ by submodularity, and $f((S\setminus{S'}) \mid S' \cup [p+1:n]) \leq f(S \setminus{S'} \mid S')$ by \eqref{eqref: conditioning}. Since the sum of these two expressions is zero, each of them must be equal to zero. In particular, $f(S') + f([p+1:n]) = f(S' \cup [p+1:n])$. As \eqref{eq: shis_18} holds for any $S \subseteq [1:p]$ such that $|S| = k$, and that $S' \subseteq S$ is arbitrary, this establishes that for all $S \subseteq [1:p]$ such that $|S| \leq k$,
\begin{align}\label{eq: shis_21}
    f(S \cup [p+1:n]) = f(S) + f([p+1:n]).
\end{align}

Let us now analyze equality in \eqref{eq: shis_6}. It is clear that equality in \eqref{eq: shis_6} holds if and only if equality holds in \eqref{eq: shis_8}. As we will see, it is sufficient to argue equality for the case $m = k-1$ in \eqref{eq: shis_8}, i.e.,
\begin{align}\label{eq: shis_19}
    \sum\limits_{\substack{S \subseteq [1:p]:\\|S| = k}} \frac{f(S \mid [p+1:n])}{\binom{p}{k}k} = \sum\limits_{\substack{S \subseteq [1:p]:\\|S| = j}} \frac{f(S \mid [p+1:n])}{\binom{p}{k-1}(k-1)}.
\end{align}
From \cite[Theorem~6.1]{Han78}\footnote{Note that \cite[Theorem~6.1]{Han78} is about entropy and thus the assertion is that any $k$ random variables from $X_1, \ldots, X_n$ are mutually independent. However, like how the underlying subset-averaging principle dictates inequality \eqref{eq: shis_8} for any submodular function $f: 2^{[1:n]} \rightarrow \mathbb{R}$, the underlying equality condition is that $f(S)=\sum_{i\in S}f(\{i\})$, for all $S\subseteq [1:n]$ such that $|S| = k$, which on substituting $f(S) = H(X_S)$ yields mutual independence of any $k$ random variables. Further, this assertion is applied to the contracted submodular set function $g: S\subseteq [1:p] \mapsto f(S \mid [p+1:n])$ in \eqref{eq: shis_20}.}, equality in \eqref{eq: shis_19} holds if and only if $\forall S \subseteq [1:p]$ such that $|S| = k$,
\begin{align}\label{eq: shis_20}
    f(S \mid [p+1:n]) = \sum\limits_{i \in S} f(\{i\} \mid [p+1:n]).
\end{align}
Combining \eqref{eq: shis_20} with the assertion in \eqref{eq: shis_21}, we get
\begin{align}\label{eq: shis_22}
    f(S) = \sum\limits_{i \in S} f(\{i\}),~\forall S \subseteq [1:p]: |S| = k.
\end{align}
We will now show that \eqref{eq: shis_22} holds $\forall S \subseteq [1:p]: |S| \leq k$. For any $S \subseteq [1:p]: |S| = k$, and $S' \subseteq S$, \eqref{eq: shis_22} implies that
\begin{align}
    0 & = \sum\limits_{i \in S} f(\{i\}) - f(S) \label{eq: shis_25}\\
    & = \sum\limits_{i \in S'} f(\{i\}) + \sum\limits_{i \in S \setminus{S'}} f(\{i\}) - f(S') - f(S \setminus{S'} \mid S') \label{eq: shis_23}\\
    & = \underbrace{\sum\limits_{i \in S'} f(\{i\}) - f(S')}_{\geq 0} + \underbrace{\sum\limits_{i \in S \setminus{S'}} f(\{i\}) - f(S \setminus{S'} \mid S')}_{\geq 0}, \label{eq: shis_24}
\end{align}
where \eqref{eq: shis_23} follows from chain rule~\eqref{eqref: chain_rule}. Notice that both the expressions in \eqref{eq: shis_24} are non-negative because $f(S') \leq \sum\limits_{i \in S'} f(\{i\})$ by submodularity, and $f(S \setminus{S'} \mid S') \leq f(S \setminus{S'}) \leq \sum\limits_{i \in S \setminus{S'}} f(\{i\})$ by \eqref{eqref: conditioning} and submodularity. Since the sum of these two expressions is zero, each of them must be equal to zero. In particular, $f(S') = \sum\limits_{i \in S'} f(\{i\})$. As \eqref{eq: shis_25} holds for any $S \subseteq [1:p]$ such that $|S| = k$, and that $S' \subseteq S$ is arbitrary, this establishes that
\begin{align}\label{eq: shis_26}
    f(S) = \sum\limits_{i \in S} f(\{i\}),~\forall S \subseteq [1:p]: |S| \leq k.
\end{align}

Finally, we are left with analyzing equality in \eqref{eq: shis_11}. It is immediate from the application of \cite[Theorem~4]{JakharKCP25} to the contracted submodular function $g: S\subseteq [p+1:n] \mapsto f(S)$ that $f(S)=\sum_{i\in S}f(\{i\})$, for all $S\subseteq [p+1:n]$. This concludes the proof of Theorem~\ref{StrongHanInequalitySubmodular}.

\section{Proof of Proposition~\ref{StrongHanInequality}}\label{appendix: c}
Since differential entropy is a submodular function, we invoke Theorem~\ref{StrongHanInequalitySubmodular} with $f(S) = h(X_S),~S \subseteq [1:n]$, and obtain
\begin{align}
    h(X_{[1:n]}) &\leq \frac{1}{\binom{p-1}{k-1}} \sum\limits_{\substack{S \subseteq [1:p]: \\ |S| = k}} h(X_{S}| X_{[p+1:n]}) + h(X_{[p+1:n]}) \label{eq: StrongHanE1}\\
    & \leq \frac{1}{\binom{n-1}{k-1} }  \sum\limits_{\substack{S \subseteq [1:n]: \\ |S| = k}} h(X_S). \label{eq: StrongHanE2}
\end{align}
It is clear that equality is trivial in \eqref{eq: StrongHanE1} for $k = p$. The equality conditions for \eqref{eq: StrongHan} imply that for $k < p$, equality in \eqref{eq: StrongHanE1} holds if and only if
\begin{align}
    h(X_S|X_{[p+1:n]}) = \sum\limits_{i \in S} h(X_i|X_{[p+1:n]}),~\forall S \subseteq [1:p],
\end{align}
which further holds if and only if $X_1, \ldots, X_p$ are conditionally independent given $X_{[p+1:n]}$. The equality conditions for \eqref{eq: StrongHan1} imply that equality in \eqref{eq: StrongHanE2} holds if and only if
\begin{enumerate}[label=(\roman*), leftmargin=*, align=left, widest=iii, nosep]
    \item for all $S \subseteq [1:p]$ such that $|S| \leq k,~h(X_S) = \sum_{i \in S} h(X_{i})$, which is equivalent to any $k$ random variables taken from $\{X_1, \ldots, X_p\}$ being mutually independent,
    \item for all $S \subseteq [1:p]$ such that $|S| \leq k,~h(X_S,X_{[p+1:n]}) = h(X_S) +  h(X_{[p+1:n]})$, which is equivalent to any $k$ random variables taken from $\{X_1, \ldots, X_p\}$ being independent of $X_{[p+1:n]}$, and
    \item for all $S \subseteq [p+1:n],~h(X_S) = \sum_{i \in S} h(X_{i})$, which is equivalent to $X_{p+1},\dots,X_n$ being mutually independent.
\end{enumerate}

\section{Details of Remark~\ref{HanStrongerThanSubadd}}\label{appendix: b}
The remark follows from the monotonically decreasing nature of the sequence of average of differential entropies over all equi-sized subsets of random variables $X_1, \ldots, X_n$ ~\cite{DemboCT91,Han78}, in particular,
\begin{align}
    \sum\limits_{S \subseteq [1:p]: |S| = k} \frac{h(X_S|X_{[p+1:n]})}{\binom{p}{k}k} \leq \sum\limits_{i = 1}^p \frac{h(X_i|X_{[p+1:n]})}{p}.
\end{align}

\section{Proof of Theorem~\ref{ConditionalSubadditivitySubmodular}}\label{appendix: d}
We first show why the inequality \eqref{eq: conditionalSub2} holds.
\begin{align}
    &\sum\limits_{S \in \mathcal{P}} f(S) \nonumber \\
    & = \sum\limits_{S \in \mathcal{P}} \big( f(S \cap [p+1:n]) + f(S \cap [1:p] \mid S \cap [p+1:n]) \big) \label{eq: csub_1}\\
    & \geq \sum\limits_{S \in \mathcal{P}} \big( f(S \cap [p+1:n]) + f(S \cap [1:p] \mid [p+1:n]) \big) \label{eq: csub_2}\\
    & = \sum\limits_{S \in \mathcal{P}} f(S \cap [p+1:n]) + \sum\limits_{S \in \mathcal{P}} f(S \cap [1:p] \mid [p+1:n]) \\
    & \geq f([p+1:n]) + \sum\limits_{S \in \mathcal{P}} f(S \cap [1:p] \mid [p+1:n]) \label{eq: csub_3}\\
    & = \sum\limits_{S \in \mathcal{P}'} f(S \mid [p+1:n]) + f([p+1:n]), \label{eq: csub_4}
\end{align}
where \eqref{eq: csub_1} holds by chain rule, \eqref{eq: csub_2} follows from \eqref{eqref: conditioning}, and \eqref{eq: csub_3} follows because $f([p+1:n]) \leq \sum\limits_{S \in \mathcal{P}} f(S \cap [p+1:n])$ by submodularity. Putting together  \eqref{eq: csub_4}, and partition subadditivity \eqref{eq: PartitionSubadditivity} applied to the contracted submodular set function $g: S\subseteq [1:p] \mapsto f(S \mid [p+1:n])$, i.e.,
\begin{align}\label{eq: csub_7}
    f([1:p]\mid [p+1:n]) \leq \sum\limits_{S \in \mathcal{P}'} f(S \mid [p+1:n]),
\end{align}
we get
\begin{align}
    f([1:n]) &\leq \sum\limits_{S \in \mathcal{P}'} f(S \mid [p+1:n]) + f([p+1:n]) \label{eq: csub_5} \\
    & \leq \sum\limits_{S \in \mathcal{P}} f(S). \label{eq: csub_6}
\end{align}

Let us now analyze the equality conditions in \eqref{eq: csub_5} and \eqref{eq: csub_6}. It follows that equality in \eqref{eq: csub_5} holds if and only if equality in \eqref{eq: csub_7} holds, i.e., $f([1:p]\mid [p+1:n]) = \sum\limits_{S \in \mathcal{P}'} f(S \mid [p+1:n])$. Clearly, equality in \eqref{eq: csub_6} holds if and only if equality is attained in each of \eqref{eq: csub_2}, and \eqref{eq: csub_3}. The equality in \eqref{eq: csub_2} is immediate if and only if, $\forall S \in \mathcal{P}$,
\begin{align}
    f(S \cap [1:p] \mid S \cap [p+1:n]) = f(S \cap [1:p] \mid [p+1:n]),
\end{align}
which further implies that, $\forall S \in \mathcal{P}$,
\begin{align}
    f(S) = f(S \cap [p+1:n]) + f(S \cap [1:p] \mid [p+1:n])
\end{align}
by application of chain rule~\eqref{eqref: chain_rule}. It follows that equality in \eqref{eq: csub_3} holds if and only if $f([p+1:n]) = \sum\limits_{S \in \mathcal{P}} f(S \cap [p+1:n])$. 

\section{Proof of Proposition~\ref{ConditionalSubadditivity}}\label{appendix: e}
Since differential entropy is a submodular function, we invoke Theorem~\ref{StrongHanInequalitySubmodular} with $f(S) = h(X_S),~S \subseteq [1:n]$, and obtain
\begin{align}
    h(X_{[1:n]}) &\leq \sum\limits_{S \in \mathcal{P}'} h(X_S|X_{[p+1:n]}) + h(X_{[p+1:n]}) \label{eq: conditionalSubE1} \\
    & \leq \sum\limits_{S \in \mathcal{P}} h(X_S). \label{eq: conditionalSubE2}
\end{align}
The equality conditions for \eqref{eq: conditionalSub1} imply that equality in \eqref{eq: conditionalSubE1} holds if and only if $h(X_{[1:p]}|X_{[p+1:n]}) = \sum_{S \in \mathcal{P}} h(X_{S \cap [1:p]}|X_{[p+1:n]})$, or equivalently, $X_{S}$, for $S\in\mathcal{P}'$ are conditionally independent given $X_{[p+1:n]}$, and the equality conditions for \eqref{eq: conditionalSub2} imply that equality in \eqref{eq: conditionalSubE2} holds if and only if
\begin{enumerate}[label=(\roman*), leftmargin=*, align=left, widest=ii, nosep]
    \item $X_{S \cap [1:p]}$ and $X_{S^{\mathrm{c}}\cap[p+1:n]}$ are conditionally independent given $X_{S\cap[p+1:n]}$, for all $S\in\mathcal{P}$, and
    \item $X_{S\cap [p+1:n]}$, for $S\in\mathcal{P}$ are independent.
\end{enumerate}

\section{Proof of Theorem~\ref{StrongSzasz}}\label{appendix: f}
Let $(X_1, \ldots, X_n)$ be a zero-mean Gaussian random vector with covariance matrix $K$, then
\begin{align}\label{eq: StrongSzasz_1}
    h(X_{[1:n]}) = \frac{1}{2}\ln{(2\pi e)^n |K|},
\end{align}
\begin{align}\label{eq: StrongSzasz_2}
    h(X_{P^{\mathrm{c}}}) = \frac{1}{2}\ln{(2\pi e)^{n-p} |K(P^{\mathrm{c}})|},
\end{align}
and $\forall S \subseteq P$ such that $|S| = k$,
\begin{align}
   &h(X_{S}|X_{P^{\mathrm{c}}}) \nonumber \\
   &= h(X_{S \cup P^{\mathrm{c}}}) -  h(X_{P^{\mathrm{c}}}) \\
   &= \frac{1}{2}\ln{\left((2\pi e)^{k+n-p} |K(S \cup P^{\mathrm{c}})|\right)} - \frac{1}{2}\ln{\left(2\pi e)^{n-p} |K(P^{\mathrm{c}})|\right)} \\
   & = \frac{1}{2} \ln{\left((2\pi e)^k \frac{|K(S \cup P^{\mathrm{c}})|}{|K(P^{\mathrm{c}})|}\right)}. \label{eq: StrongSzasz_3}
\end{align}
We now invoke Proposition~\ref{StrongHanInequality}
\begin{align}
    h(X_{[1:n]}) &\leq \frac{1}{\binom{p-1}{k-1}} \sum\limits_{\substack{S \subseteq P: \\ |S| = k}} h(X_{S}|X_{P^{\mathrm{c}}}) + h(X_{P^{\mathrm{c}}}) \label{eq: StrongSzasz_4} \\
    & \leq \frac{1}{\binom{n-1}{k-1} }  \sum\limits_{\substack{S \subseteq [1:n]: \\ |S| = k}} h(X_S), \label{eq: StrongSzasz_7}
\end{align}
and substitute the entropy values from \eqref{eq: StrongSzasz_1}, \eqref{eq: StrongSzasz_2}, and \eqref{eq: StrongSzasz_3} in \eqref{eq: StrongSzasz_4} to obtain
\begin{align}
    \frac{1}{2}\ln{(2\pi e)^n |K|} &\leq \frac{1}{\binom{p-1}{k-1}} \sum\limits_{\substack{S \subseteq P: \\ |S| = k}} \frac{1}{2} \ln{\left((2\pi e)^k \frac{|K(S \cup P^{\mathrm{c}})|}{|K(P^{\mathrm{c}})|}\right)} \nonumber \\
    &\quad\quad + \frac{1}{2}\ln{(2\pi e)^{n-p} |K(P^{\mathrm{c}})|},
\end{align}
which further implies that
\begin{align}\label{eq: StrongSzasz_6}
    |K| \leq |K(P^{\mathrm{c}})|\left( \prod\limits_{\substack{S \subseteq P:\\|S|=k}} \frac{|K(S \cup P^{\mathrm{c}})|}{|K(P^{\mathrm{c}})|}\right)^{\frac{1}{\binom{p-1}{k-1}}}.
\end{align}
Further, substituting
\begin{align}
    h(X_S) = \frac{1}{2}\ln{(2\pi e)^{|S|} |K(S)|},~\forall S \subseteq [1:n]: |S| = k,
\end{align}
in \eqref{eq: StrongSzasz_7} gives
\begin{align}\label{eq: StrongSzasz_5}
    |K(P^{\mathrm{c}})|\left( \prod\limits_{\substack{S \subseteq P:\\|S|=k}} \hspace{-0.3em}\frac{|K(S \cup P^{\mathrm{c}})|}{|K(P^{\mathrm{c}})|}\right)^{\frac{1}{\binom{p-1}{k-1}}} \hspace{-0.4em}\leq \left(\prod\limits_{\substack{S \subseteq [1:n]:\\|S|=k}} \hspace{-0.5em}|K(S)|\right)^{\frac{1}{\binom{n-1}{k-1}}}\hspace{-0.55em}.
\end{align}

Notice that the equality conditions for \eqref{eq: StrongSzasz_7} imply pairwise independence of random variables $X_1, \ldots, X_n$ for equality to hold in \eqref{eq: StrongSzasz_5} when $2 \leq k \leq p$. This gives us that $K$ has to be diagonal for equality to hold in \eqref{eq: StrongSzasz_5}. Thus, the bound on |K| in \eqref{eq: SS_1} is always strictly less than the bound obtained from Sz\'asz inequality, when $K$ is not a diagonal matrix and $2 \leq k \leq p$.

In contrast, for $k = 1$ and $p < n$, the equality conditions for \eqref{eq: StrongSzasz_7} imply only the pairwise independence of random variables $X_{p+1}, \ldots, X_n$ and pairwise independence of $X_i$ and $X_j,~i \in P, j \in P^{\mathrm{c}}$, which further implies that $K(P^{\mathrm{c}})$ is diagonal and $K(P,P^{\mathrm{c}}) = 0$, respectively.

Let us now analyze the equality conditions in \eqref{eq: StrongSzasz_6}. For $k = p$, there is nothing to analyze. For $k < p$, it is clear that equality holds in \eqref{eq: StrongSzasz_6} if and only if equality is attained in \eqref{eq: StrongSzasz_4}, which further holds if and only if the random variables $X_1, \ldots, X_p$ are conditionally independent given $X_{[p+1:n]}$. Since, mutual independence implies pairwise independence for random variables, $X_1, \ldots, X_p$ being conditionally independent given $X_{}$ is equivalent to the conditional covariance matrix of $X_P$ given $X_{P^{\mathrm{c}}}$ being diagonal. Using the fact that the conditional covariance matrix of a random vector $Y$ given random vector $Z$ for a jointly Gaussian random vector $(Y,Z)$ with covariance matrix $M$ is given by $M(Y) - M(Y.Z)M(Z)^{-1}M(ZY)$, we get that 
\begin{align}
    K(P) - K(P,P^{\mathrm{c}}) \big(K(P^{\mathrm{c}})\big)^{-1} K(P^{\mathrm{c}},P)
\end{align}
is diagonal.

\section{Details of Remark~\ref{LinAlgApproach}}\label{appendix: g}
Let the matrix $K$ be partitioned as
\begin{align}\label{eq: la_1}
    K = \begin{pmatrix}
        K(P) & K(P,P^{\mathrm{c}}) \\
        K(P^{\mathrm{c}},P) & K(P^{\mathrm{c}})
    \end{pmatrix},
\end{align}
and let $M$ denote the Schur complement of $K(P^{\mathrm{c}})$ in $K$, i.e.,
\begin{align}\label{eq: la_2}
    M = K(P) - K(P,P^{\mathrm{c}}) \big(K(P^{\mathrm{c}})\big)^{-1} K(P^{\mathrm{c}},P).
\end{align}
Using the fact that Schur complement of a positive definite matrix $K$ is also positive definite and that the determinant of Schur complement of $K(P^{\mathrm{c}})$ in $K$ is $|K|/|K(P^{\mathrm{c}})|$, we apply Sz\'asz inequality to the matrix $M$ to obtain
\begin{align}
    |K| &\leq |K(P^{\mathrm{c}})| \left(\prod\limits_{S \subseteq P:|S|=k} |M(S)|\right)^{\frac{1}{\binom{p-1}{k-1}}} \\
    &= |K(P^{\mathrm{c}})| \left( \prod\limits_{S \subseteq P: |S|=k} \frac{|K(S \cup P^{\mathrm{c}})|}{|K(P^{\mathrm{c}})|}\right)^{\frac{1}{\binom{p-1}{k-1}}} \label{eq: la_3},
\end{align}
where \eqref{eq: la_3} follows because $\forall S \subseteq P$ such that $|S| = k$, $M(S) = K(S) - K(S,P^{\mathrm{c}})K(P^{\mathrm{c}})^{-1}K(P^{\mathrm{c}},S)$, i.e., the Schur complement of $K(P^{\mathrm{c}})$ in $K(S \cup P^{\mathrm{c}})$.

\section{Proof of Theorem~\ref{EigenvaluesStrongKyFan}}\label{appendix: h}
Let $K'$ denote the principal submatrix corresponding to the set of indices $[1:h] \cup [p+1:n]$, and let $\lambda_1 \leq \lambda_2 \leq \ldots \leq \lambda_{h+n-p}$ and $\lambda_1' \leq \lambda_2' \leq \ldots \leq \lambda_{h+n-p}'$ denote the eigenvalues of $K$, and $K'$, respectively. Since the eigenvalues of any principal submatrix interlace the eigenvalues of the positive definite matrix it is obtained from (\!\!~\cite[Theorem~4.3.28]{HornJ12}), we have that
\begin{align}\label{eq: ESKF_1}
    \prod\limits_{i=1}^{h+\ell} \lambda_i \leq \prod\limits_{i=1}^{h+\ell} \lambda'_i.
\end{align}
Further, the determinant of the matrix $(<He_i,e_j>)_{i,j \in [1:q]}$, for a positive definite matrix $H$ and a set of orthonormal vectors $\{e_i\}_{i = 1}^q$, cannot exceed the product of the largest $q$ eigenvalues of $H$ by \cite[Lemma~3]{Fan50}. So, using this fact for positive definite $K'$ with $e_i$'s as the standard unit vectors corresponding to set of columns $[h+\ell+1:h+n-p]$ in $K'$, we have that
\begin{align}
    \lambda'_{h+\ell+1}\lambda'_{h+\ell+2}\ldots\lambda'_{h+n-p} &\geq |K'([h+\ell+1:h+n-p])| \\
    & = |K([p+\ell+1:n])|.\label{eq: ESKF_2}
\end{align}
Also, from Fischer inequality \cite[Theorem~7.8.5]{HornJ12},
\begin{align}
    |K(P^{\mathrm{c}})| &\leq |K(Q_{\ell})||K([p+\ell+1:n])|,
\end{align}
which further implies that
\begin{align}
    \frac{|K(P^{\mathrm{c}})|}{|K(Q_{\ell})|} \leq |K([p+\ell+1:n])|.\label{eq: ESKF_3} 
\end{align}
Combining \eqref{eq: ESKF_2} and \eqref{eq: ESKF_3}, we get
\begin{align}
    \lambda'_{h+\ell+1}\lambda'_{h+\ell+2}&\ldots\lambda'_{h+n-p} \geq \frac{|K(P^{\mathrm{c}})|}{|K(Q_{\ell})|}. \label{eq: ESKF_6}
\end{align}
Then, \eqref{eq: ESKF_6} and the fact that $|K'| = \prod_{i = 1}^{h+n-p} \lambda'_i$, gives
\begin{align}
    \lambda'_{1}\ldots\lambda'_{h+\ell} &\leq |K(Q_{\ell})| \frac{|K'|}{|K(P^{\mathrm{c}})|}. \label{eq: ESKF_4}
\end{align}
Applying Theorem~\ref{StrongSzasz} to positive definite $K'$, we get
\begin{align}
    &\frac{|K'|}{|K'([h+1:h+n-p])|} \nonumber \\
    &\quad \leq \left( \prod\limits_{S \subseteq [1:h]: |S|=k} \frac{|K'(S \cup [h+1:h+n-p])|}{|K'([h+1:h+n-p])|}\right)^{\frac{1}{\binom{h-1}{k-1}}},
\end{align}
which further implies that
\begin{align}
    \frac{|K'|}{|K(P^{\mathrm{c}})|} \leq \left( \prod\limits_{S \subseteq [1:h]: |S|=k} \frac{|K(S \cup P^{\mathrm{c}})|}{|K(P^{\mathrm{c}})|}\right)^{\frac{1}{\binom{h-1}{k-1}}}, \label{eq: ESKF_5}
\end{align}
as $K'([h+1:h+n-p]) = K(P^{\mathrm{c}}),$ and $K'(S \cup [h+1:h+n-p]) = K(S \cup P^{\mathrm{c}}),~\forall S \subseteq [1:h]$ such that $|S| = k$. Putting \eqref{eq: ESKF_1}, \eqref{eq: ESKF_4} and \eqref{eq: ESKF_5} together, we get
\begin{align}
    \prod\limits_{i=1}^{h+\ell} \lambda_i \leq |K(Q_{\ell})| \left( \prod\limits_{S \subseteq [1:h]: |S|=k} \frac{|K(S \cup P^{\mathrm{c}}|}{|K(P^{\mathrm{c}})|}\right)^{\frac{1}{\binom{h-1}{k-1}}}.
\end{align}

\section{Proof of Theorem~\ref{StrongFischer}}\label{appendix: i}
Let $(X_1, \ldots, X_n)$ be a zero-mean Gaussian random vector with covariance matrix $K$, then
\begin{align}\label{eq: StrongFischer_1}
    h(X_{[1:n]}) = \frac{1}{2}\ln{(2\pi e)^n |K|},
\end{align}
\begin{align}\label{eq: StrongFischer_2}
    h(X_{P^{\mathrm{c}}}) = \frac{1}{2}\ln{(2\pi e)^{n-p} |K(P^{\mathrm{c}})|},
\end{align}
and $\forall S \in \mathcal{P}'$,
\begin{align}
   &h(X_{S}|X_{P^{\mathrm{c}}}) \nonumber \\
   &= h(X_{S \cup P^{\mathrm{c}}}) -  h(X_{P^{\mathrm{c}}}) \\
   &= \frac{1}{2}\ln{\left((2\pi e)^{k+n-p} |K(S \cup P^{\mathrm{c}})|\right)} - \frac{1}{2}\ln{\left(2\pi e)^{n-p} |K(P^{\mathrm{c}})|\right)} \\
   & = \frac{1}{2} \ln{\left((2\pi e)^k \frac{|K(S \cup P^{\mathrm{c}})|}{|K(P^{\mathrm{c}})|}\right)}. \label{eq: StrongFischer_3}
\end{align}
We now invoke Proposition~\ref{ConditionalSubadditivity}
\begin{align}
    h(X_{[1:n]}) &\leq \sum\limits_{S \in \mathcal{P}'} h(X_S|X_{P^{\mathrm{c}}}) + h(X_{P^{\mathrm{c}}}) \label{eq: StrongFischer_4} \\
    & \leq \sum\limits_{S \in \mathcal{P}} h(X_S), \label{eq: StrongFischer_5}
\end{align}
and substitute the entropy values from \eqref{eq: StrongFischer_1}, \eqref{eq: StrongFischer_2}, and \eqref{eq: StrongFischer_3} in \eqref{eq: StrongFischer_4} to obtain
\begin{align}
    \frac{1}{2}\ln{(2\pi e)^n |K|} &\leq \sum\limits_{S \in \mathcal{P}'} \frac{1}{2} \ln{\left((2\pi e)^k \frac{|K(S \cup P^{\mathrm{c}})|}{|K(P^{\mathrm{c}})|}\right)} \nonumber \\
    & \quad \quad + \frac{1}{2}\ln{(2\pi e)^{n-p} |K(P^{\mathrm{c}})|},
\end{align}
which further implies that
\begin{align}\label{eq: StrongFischer_6}
    |K| \leq |K(P^{\mathrm{c}})| \prod\limits_{S \in \mathcal{P}'} \frac{|K(S \cup P^{\mathrm{c}})|}{|K(P^{\mathrm{c}})|}.
\end{align}
Further, substituting
\begin{align}
    h(X_S) = \frac{1}{2}\ln{(2\pi e)^{|S|} |K(S)|},~\forall S \in \mathcal{P},
\end{align}
in \eqref{eq: StrongFischer_5} gives
\begin{align}\label{eq: StrongFischer_7}
    |K(P^{\mathrm{c}})| \prod\limits_{S \in \mathcal{P}'} \frac{|K(S \cup P^{\mathrm{c}})|}{|K(P^{\mathrm{c}})|} \leq \prod\limits_{S \in \mathcal{P}} |K(S)|.
\end{align}
Let us analyze the equality conditions in \eqref{eq: SF_1} and \eqref{eq: SF_2}. Equality in \eqref{eq: SF_1} is the same as equality in \eqref{eq: StrongFischer_6}, and equality in \eqref{eq: StrongFischer_6} holds if and only if equality is attained in \eqref{eq: StrongFischer_4}, which further holds if and only if $X_S,~\forall S \in \mathcal{P}'$ are conditionally independent given $X_{P^{\mathrm{c}}}$. Further, this means that the submatrix corresponding to any $S, S' \in \mathcal{P}'$ such that $S \neq S'$ in the conditional covariance matrix $M$ of random vector $X_{P}$ given $X_{P^\mathrm{c}}$, $M(S,S')$ is $0$, where $M$ is given by
\begin{align}
    M = K(P) - K(P,P^{\mathrm{c}}) K(P^{\mathrm{c}})^{-1} K(P^{\mathrm{c}},P),
\end{align}
which is the Schur complement of $K(P^{\mathrm{c}})$ in $K$. 

Clearly, equality in \eqref{eq: SF_2} is the same as equality in \eqref{eq: StrongFischer_7}, and equality in \eqref{eq: StrongFischer_7} holds if and only if equality is attained in \eqref{eq: StrongFischer_5}, which further holds if and only if
\begin{enumerate}[label=(\roman*), leftmargin=*, align=left, widest=ii, nosep]
    \item $X_{S \cap P}$ and $X_{S^{\mathrm{c}}\cap P^{\mathrm{c}}}$ are conditionally independent given $X_{S\cap P^{\mathrm{c}}}$, for all $S\in\mathcal{P}$, and
    \item $X_{S\cap P^{\mathrm{c}}}$, for $S\in\mathcal{P}$ are independent.
\end{enumerate}
Equality condition (i) is equivalent to the submatrix corresponding to the set of rows $A$ and the set of columns $C$ in the conditional covariance matrix $M$ of random vector $X_{A \cup C}$ given $X_{B}$, $M(A,C)$ being $0$ for all $S \in \mathcal{P}$, where $M$ is given by
\begin{align}
    M = K(A \cup C) - K(A \cup C,B) K(B)^{-1} K(B,A \cup C).
\end{align}
Further, $M(A,C) = 0$, for all $S \in \mathcal{P}$ is equivalent to
\begin{align}
    K(A,C) = K(A,B)K(B)^{-1}K(B,C),~\forall S \in \mathcal{P}.
\end{align}
Equality condition (ii) is equivalent to the submatrix corresponding to any $S \cap P^{\mathrm{c}}, S' \cap P^{\mathrm{c}}$ in the covariance matrix $K$, $K(S \cap P^{\mathrm{c}}, S' \cap P^{\mathrm{c}})$ being $0$,~$\forall S,S' \in \mathcal{P}$ such that $S \neq S'$.

\fi
\end{document}